# Comparing how Large Language Models perform against keyword-based searches for social science research data discovery

Mark Green[1], Maura Halstead[2], Caroline Jay[2], Richard Kingston[3], Alex Singleton[1], David Topping[4].

1 Department of Geography and Planning, University of Liverpool, Liverpool, UK

2 Department of Computer Science, School of Engineering, University of Manchester, Manchester, UK

3 Department of Planning, Property and Environmental Management, University of Manchester, Manchester, UK

4 Department of Earth and Environmental Sciences, University of Manchester, Manchester, UK

## Introduction

The analysis presented here compares the search results generated from the large language model (LLM) derived semantic search tool compared to the equivalent existing keyword search tool. The aim of the analysis was to evaluate whether the semantic search tool produced similar results to the keyword search tool. We test the following hypotheses:

1. The LLM semantic search tool will give a larger number of search results than compared to the keyword-based search.
2. The LLM semantic search tool captures all the results found in the keyword-based search.
3. The most relevant search results (i.e., first 10 results) are different on both search tools since they capture and prioritise different issues.
4. The LLM semantic search can handle erroneous, obscure or complex queries better than the keyword search.

In brief, we compared the following two data discovery tools:

1. We developed our own bespoke semantic search tool which used Large Language Models to process metadata records for data across several UKRI data services. The resulting tool matches up the semantic meaning of a user's search query to the semantic database of metadata. You can find the tool here https://apps.geods.ac.uk/semantic-catalogue/Search. We have detailed the methods in depth here https://github.com/GeographicDataService/metadata-semantic-search.

2. We used the Consumer Data Research Centre (CDRC) website and their keyword-based traditional search box. The CDRC website is available here https://www.cdrc.ac.uk/, although the service has closed since we completed this analysis.

**Methodology**

Data

We accessed the search logs from the Consumer Data Research Centre (CDRC) website (specifically http://data.cdrc.ac.uk) for the time period 2023-12-07 15:32:42 to 2024-10-31 11:04:35. This allowed us to evaluate how research users search for datasets to provide a more realistic real-world analysis. We elected to focus on the most common searches during the time period. To identify these, we calculated the frequency of each search term over this period. We then selected the top 200 most frequently searched terms (each had a minimum of 35 searches). Within the top 200 terms, there were several instances of search terms which were produced by computer interactions (e.g., 'sites default files x', 'csd security', 'phpmyadmin', 'gateway login'). We removed all these terms leaving an analytical sample size of n = 131 (65.5%). Table 1 presents the most frequent search terms.

**Table 1: Top 10 most frequent search terms on the Consumer Data Research Centre website.**

| Term | Rank |
|---|---|
| national | 1 |
| lad | 2 |
| imd | 3 |
| london | 4 |
| footfall | 5 |
| retail | 6 |
| deprivation | 7 |
| camden | 8 |
| transport | 9 |
| ahah | 10 |

The search terms were then run on both the CDRC website 'data' search tool (which uses a keyword based search algorithm to identify relevant datasets) and the LLM derived semantic search tool (https://apps.cdrc.ac.uk/semantic-catalogue/Search). For the former, we automated the process and scraped off all the search results of the website including number of results returned, dataset titles and descriptions. It was difficult to scrape the semantic search results due to how the website is constructed.

Here, we manually entered each term. Search results were logged via the 'langsmith' resource which underpins the keyword search tool. This produced a JSON containing all the search results returned for each query. We then filtered results for just those relating to the CDRC data catalogue so that the findings were comparable.

The semantic search tool currently, at the time of this report, only stores up to 500 search results at a time. We tested increasing this to 5000 and 10 000 to capture all search results but this caused issues for the search tool (e.g., could not store or display results as data size was too large, hourly pinecone API limits exceeded). While the 500 search results limit may produce outputs which do not capture all datasets returned (since the 500 limit is before filtering for CDRC results), our testing suggested that it captured nearly all of them. Additionally, since we designed the tool to exclude results with low relevance, the LLM tool is unlikely to return a very high number of results (e.g., it mostly returns <50 results), unlike the keyword search where all results of any relevance will be returned.

Statistical analysis

Summary statistics were calculated to describe the number of datasets each search method returned. These were supplemented by visualisations. To assess the similarity of search results, we employed a mixed-methods approach. First, we qualitatively described the key themes of datasets returned in each method. Second, we quantitatively summarise the similarity of datasets using three metrics:

1. Exact matches of datasets: The percentage of datasets returned in the LLM semantic search results which were present in the keyword search. The metric informs whether the exact same datasets are being captured. Values range from 0% (no datasets overlapping) to 100% (all of keyword returned results present in LLM semantic search results).
2. Jaccard's similarity: The index captures the 'literal' similarity of terms. Here this would evaluate whether the same exact words are found in each search results. The indicator ranges from 0 (no overlap) to 1 (complete overlap).
3. Cosine similarity: We assess the semantic similarity of terms in dataset titles. This extends the Jaccard's score since it captures words that are similar in meaning (e.g., retail and shopping are similar terms, but would not be considered this on Jaccard's similarity as they are different terms). It also accommodates for if the LLM semantic search returns a longer list of relevant results (including new datasets) since this would produce a lower Jaccard's score since there are more terms not overlapping. We used a pretrained language model (BERT) to extract the embeddings of terms within the titles of datasets returned. We then estimate the cosine similarity between these embeddings to describe how similar the terms are. The approach would capture cases where the LLM semantic search returns similar topics that are relevant, even if they are not the exact same as the datasets in the keyword search. The score ranges from a minimum of -1 (each search tool gives directly opposite terms), to 0 (no similarity) and a maximum of 1 (terms our identical in semantic meaning).

**Results**

Hypothesis 1: The LLM semantic search tool will give a larger number of search results than compared to the keyword-based search.

The median number of results for the semantic search tool was higher (median = 30, interquartile range (IQR) = 20, 41.5) than compared to the keyword search tool (median = 7, IQR = 2, 17). A Welch two sample t-test suggested that this difference was statistically significantly different (t = 5.87, p = <0.001). Figure 1 presents the association between the number of search results returned. There was little agreement between the two tools, as evidenced by a weak positive correlation estimate of r=0.18 (Pearson's coefficient). There were three noticeable outliers (i.e., search results >100) which had a larger number of results in the keyword-based search (terms = 'events', 'property' and 'data'). There were no outliers of the same scale detected for the LLM tool. This partly reflects that the LLM search tool is designed not to present results with low relevance. Removing the three keyword outliers from the data did not significantly change the t-test estimate (t = 8.67, p = <0.001), but it led to the weak positive correlation being attenuated (r = 0.03).

Figure 1 demonstrates how search queries which gave no results on the keyword-based tool had returned results for the same query on the LLM tool. 15 of the 16 occurrences where this happened involved queries of specific geographic places (e.g., Cambridge, Bradford, Sheffield). The LLM results returned included location specific datasets about a specific place that were otherwise not captured using the keyword search. The LLM tool also appears to interpret that national level datasets which have sub-regional or place-based components will contain a specific place (e.g., datasets returned included Local Data Company footfall data, broadband speed estimates, Output Area Classification which each have a spatial component). In comparison, datasets which have a wide geographical component (e.g., small area data for Great Britain) do not have every possible place tagged as a keyword.

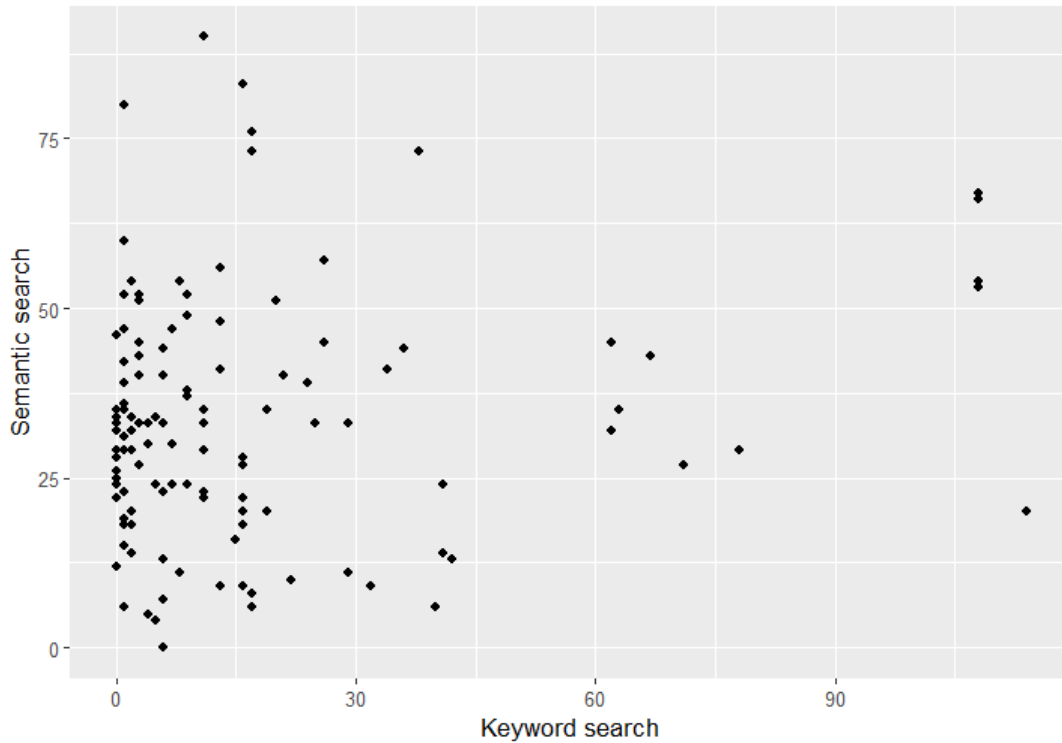

**Figure 1: Scatter plot presenting the number of results returned for each term by the keyword and semantic search tools.**

When we examined some of these queries using the other data catalogues, the LLM tool also highlighted datasets of nearby places. For example, searching for 'Sheffield' picked up datasets collected for 'Yorkshire' or 'South Yorkshire' demonstrating that it interpreted where the regional context of was for Sheffield's located. Similarly, it also returned the UKDS's dataset 'Oral histories of homes and daily lives in Stocksbridge and Stevenage' with the 'Ask AI' explaining that:

"*The dataset on oral histories from Stocksbridge and Stevenage is relevant to the query about Sheffield because Stocksbridge is a town located near Sheffield, and the oral histories may provide insights into the daily lives and experiences of residents in this region. Understanding the historical context and personal narratives from nearby areas can enhance knowledge about the cultural and social dynamics of Sheffield itself. Additionally, the dataset's focus on energy demand and infrastructure adaptations may reflect broader trends applicable to Sheffield's urban development and community practices.*"

One caveat was that further down the search results, the LLM tool started to highlight datasets that did not contain the place itself with a focus on comparative analyses. For example, the query 'Sheffield' returned the dataset "Identities and Regeneration in the Former Coalfields of East Durham". This was explained by the 'Ask AI' as:

*"The dataset on "Identities and Regeneration in the Former Coalfields of East Durham" is relevant to the query about Sheffield as both regions share a historical connection to coal mining and industrial decline. The research explores how communities in former coalfield areas, like those in East Durham, navigate identity and regeneration, which can provide insights into similar processes occurring in Sheffield, a city with its own coal mining heritage. Additionally, understanding the social dynamics and regeneration efforts in one coalfield community may offer valuable lessons for Sheffield as it addresses its own post-industrial challenges."*

There were two occurrences where the LLM search tool gave no results. This was the queries 'topic education' and 'topic about education'. Both occasions the search results were dominated by results from UKDS and no CDRC datasets were returned in the 500 results returned. Datasets were returned but these were identified as low relevance. The similar query of 'education' produced seven results on the LLM tool (compared to six on the keyword tool) suggesting that this was an outlier. This appeared to be the combinations of terms used here which had less qualitative meaning when placed together. This offers one area for further model development.

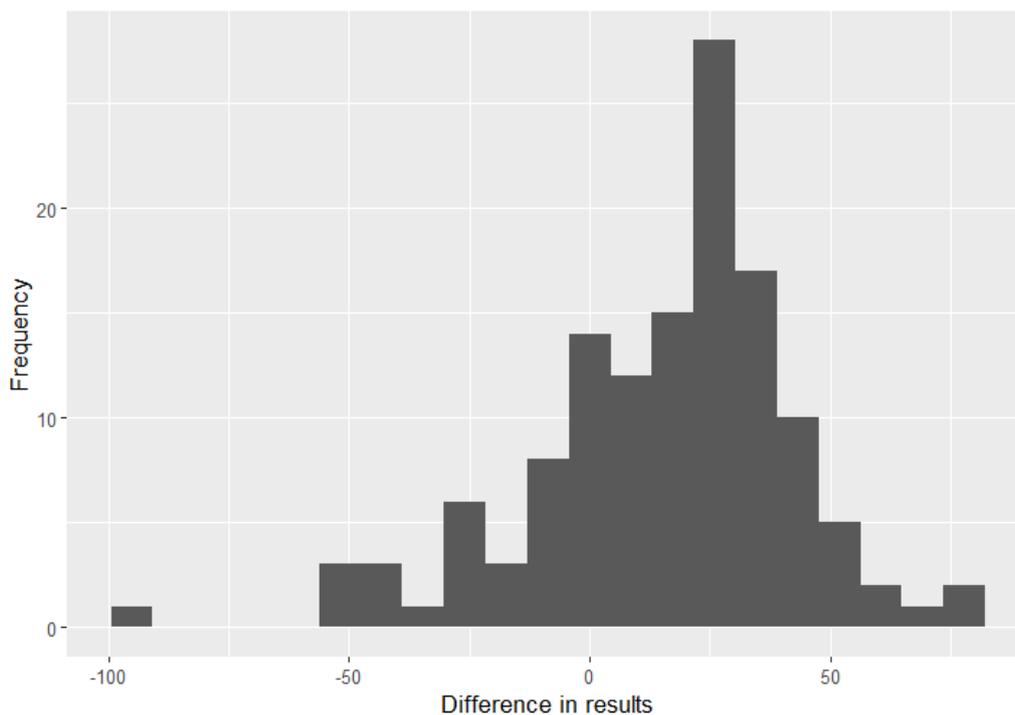

**Figure 2: A histogram presenting the difference in the number of search results returned by term (excludes outliers with values <110).**

The median difference between the two search tools was 19 results higher for the LLM semantic tool (IQR = 1, 32.5). Figure 2 presents the distribution of these differences. Most search results were more frequent in the LLM semantic tool (i.e., 79% of terms had a difference greater than 0). The scale of these differences was not always large. For example, 36% of all terms had a difference between 0 and 25 higher in the LLM semantic tool. The terms with the largest differences (defined here as a difference greater than or equal to 50) were observed for the LLM semantic search (6% of all terms were >=50) than compared to the LLM search tool (2% were <= 50).

The terms which the LLM semantic search gave a larger number of results than the keyword search tended to be in relation to specific datasets (see Table 2 for a full list). For example, the query 'pdv' relates to a single CDRC dataset titled '[PDV Consumer Lifestyle surveys](#)'. The keyword search only gives this as its single result. For the LLM, the same data is returned as the result with the highest relevance (same for other similar occurrences). The other results include a range of themes including transport, property/housing, digital and population datasets. The 'Ask AI' button does not give a clear or consistent reason behind its decisions suggesting low comprehension. This is likely reflecting that the term 'pdv' is not a term it interpreted due to being an acronym of infrequent use.

The query 'airbnb' follows a similar result for the keyword search above, only returning two results (both specific Airbnb datasets). The LLM search tool presents those same two datasets as the two top results. Unlike 'pdv', the LLM has a contextual understanding of what Airbnb means, and all the other datasets returned either are focused on housing (e.g., house prices, rental prices, house sales) or are broader datasets which contain housing measures within them. Here the LLM search tool gives a wider range of datasets that are relevant that the keyword search misses.

The query 'lad about liverpool' gave few results in the keyword search and many in the LLM search. This can be explained by the query which contains 'lad about'. Searching for 'Liverpool' in the keyword search gives 25 results. The LLM search tool can understand that the 'lad about' does not give much value and focuses on returning datasets either about Liverpool or that might have some relevance to studying Liverpool. This is an important feature as the keyword tool is spelling or term dependent, with the LLM tool more flexible.

For 'geodata', the LLM gives a larger number of datasets and captures most CDRC data that have a geographical component to them. The lower results for the keyword search appears to be based on fewer of these datasets containing the specific keyword. So it appears here that the LLM tool performs better. The results here are similar for the query 'maps' where the LLM tool gives geographical data that it explains as being able to be mapped. In contrast, the keyword search results give two datasets called Money and Pension Service (MaPS) that contain 'maps' in the title. These two surveys are not returned in the LLM search.

The query 'postcode' provides relevant results returned on the LLM tool including those datasets which include postcode information (e.g., WhenFresh/Zoopla Property Transactions, Rentals and Associated Migration), are constructed from postcode data

(e.g., access to healthy assets and hazards) or have geographical data linked at the postcode level (e.g., British Population Survey). It also gives general small area datasets which it explains as being able to be matched to postcodes for further analysis (e.g., London Output Area Classification). The keyword tool does not capture such information across all these types of results, since it relies on the LLM making contextual insights about applicability of data or knowing how a dataset was created.

The terms which the keyword search gave a larger number of results than the LLM search tool were general terms (e.g., 'census', 'data', 'events', 'taxonomy', 'population'). These terms tended to be tagged on to a lot of datasets as keywords since they were applicable, but otherwise offer little value in describing what a dataset contains. As such, the LLM gave fewer results as it could not find good matches due to the broad nature of the query.

**Table 2: Most common terms that produced the largest differences in the number of results returned by each search tool.**

| Query | Keyword | LLM | Difference |
|---|---|---|---|
| *Highest differences where LLM gave more* | | | |
| geodata | 11 | 90 | 79 |
| lad about liverpool | 1 | 80 | 79 |
| local data company | 16 | 83 | 67 |
| pdv | 1 | 60 | 59 |
| maps | 17 | 76 | 59 |
| postcode | 17 | 73 | 56 |
| airbnb | 2 | 54 | 52 |
| *Highest differences where keyword gave more* | | | |
| events | 114 | 20 | -94 |
| taxonomy | 108 | 53 | -55 |
| property | 108 | 54 | -54 |
| consumer | 78 | 29 | -49 |
| population | 71 | 27 | -44 |
| data | 108 | 67 | -41 |
| census | 40 | 6 | -34 |

Overall, we accept hypothesis 1 as it is evident that the LLM tool returns more results most of the time.

Hypothesis 2: The LLM semantic search tool captures all the results found in the keyword-based search.

To examine how similar the LLM results are to the keyword results, we calculated the percentage of all datasets in the keyword search results returned which are also returned in the LLM search results. We removed the 16 occurrences where the keyword search tool returned no results (analytical sample is therefore n=115). The median percentage of all keyword returned datasets captured in the LLM search results was 54.6% (IQR = 25.7%, 96.2%). 25.2% of the queries saw all the keyword search results returned in the LLM results, with 9.6% having none of the keyword results returned.

While the LLM results do not always return all the same datasets as found in the keyword search tool, it is plausible that the LLM viewed some of the keyword results as less relevant and therefore opted to not return them. This may further be explained if the LLM results are returning other datasets that are more relevant about the same topic. To assess this, we evaluated the semantic similarity of the datasets returned in the LLM results to those in the keyword results. We expect that the datasets in the LLM are similar in content and themes, suggesting that they give a relevant set of results.

Figures 3 and 4 present the similarity of results across the keyword and LLM search tools by Jaccard's and cosine similarity respectively. Subjective interpretation of the values can be as follows: (i) Jaccard's values >=0.3 suggest good overlap, with values 0.1 to 0.3 having some overlap, and values <0.1 indicating little to no overlap. (ii) Cosine values >= 0.7 indicate strong similarity of terms, with values <0.5 suggesting weaker similarity. One should interpret values carefully as weaker values can still be semantically similar if one is comparing longer pieces of text. There is a strong positive correlation between the two metrics (r = 0.75).

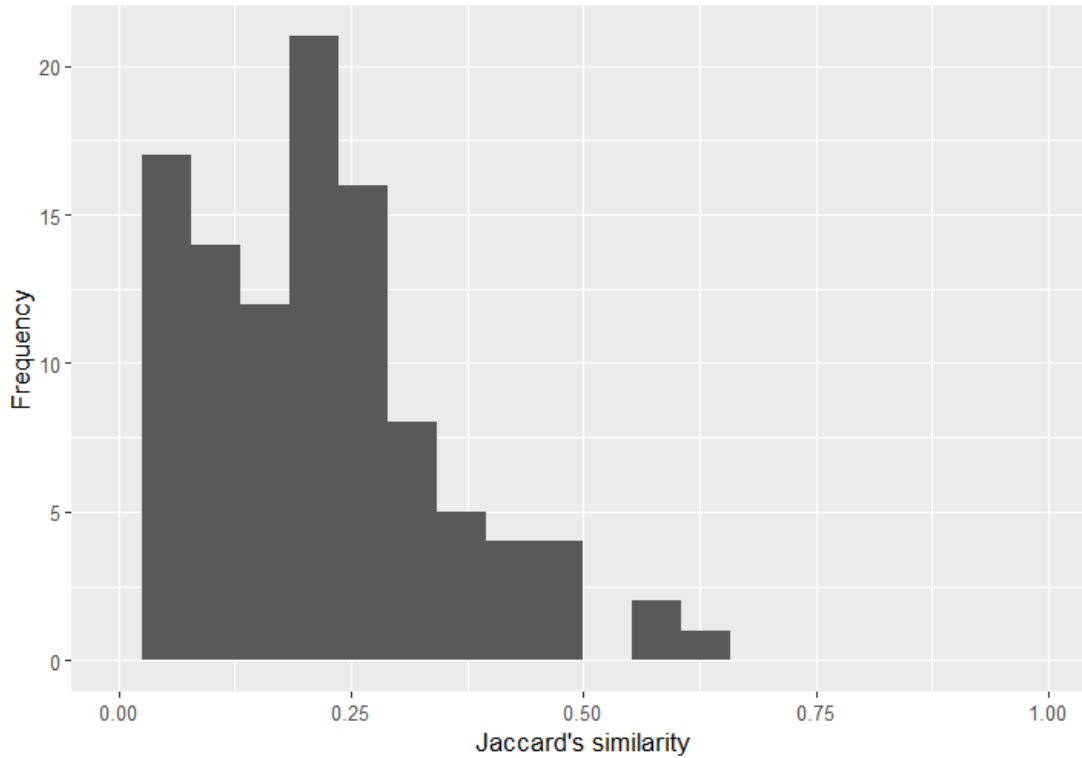

**Figure 3: A histogram presenting the Jaccard's similarity score of query search results.**

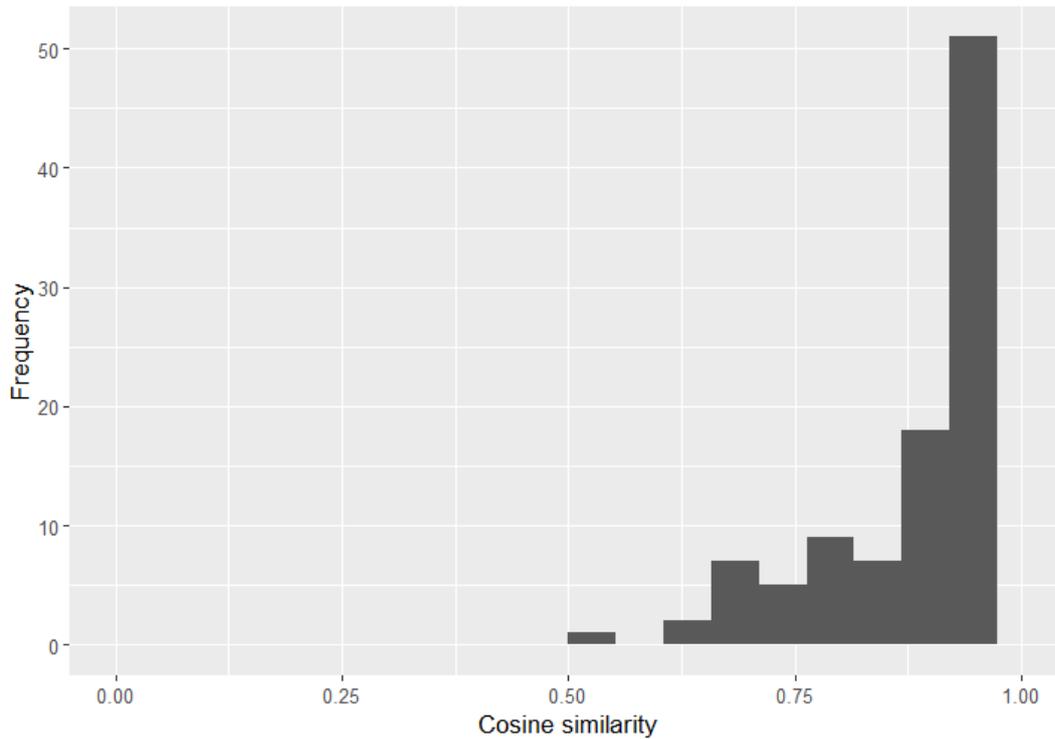

**Figure 4: A histogram presenting the cosine similarity score of query search results.**

The Jaccard's similarity distribution (Figure 3) displays a positive skew of results (median 0.20, IQR = 0.09, 0.28). It suggests that results returned in the LLM have some overlap of terms in the keyword search, but not strongly on average. 19.5% have good similarity (>=0.3) and 27.5% have little to no overlap of terms (<0.1). The cosine similarity patterns were also skewed, but towards the higher values suggesting high semantic similarity (median 0.94, IQR = 0.85, 0.96). 92% of values were greater than or equal to 0.7 (63.7% above 0.9) and none were less than 0.5. These two plots would suggest that while the LLM does not necessarily capture the exact same terms/datasets for each query, it is capturing semantically similar datasets which are relevant.

There were no distinct themes in types of queries which scored higher or lower on either index. Terms that scored lowest across the two measures included 'digital iuc', 'pdv', 'camden' and 'manchester'. Terms that scored highest included 'data', 'mobility', 'classification' and 'energy'. Examining Figure 5, we can identify some queries where the values are different to their relative position (i.e., outliers). Outliers tend to be queries where the keyword search gives a low number of results (e.g., 'surname', 'huq', 'house price' and 'loac' have 2-4 results each).

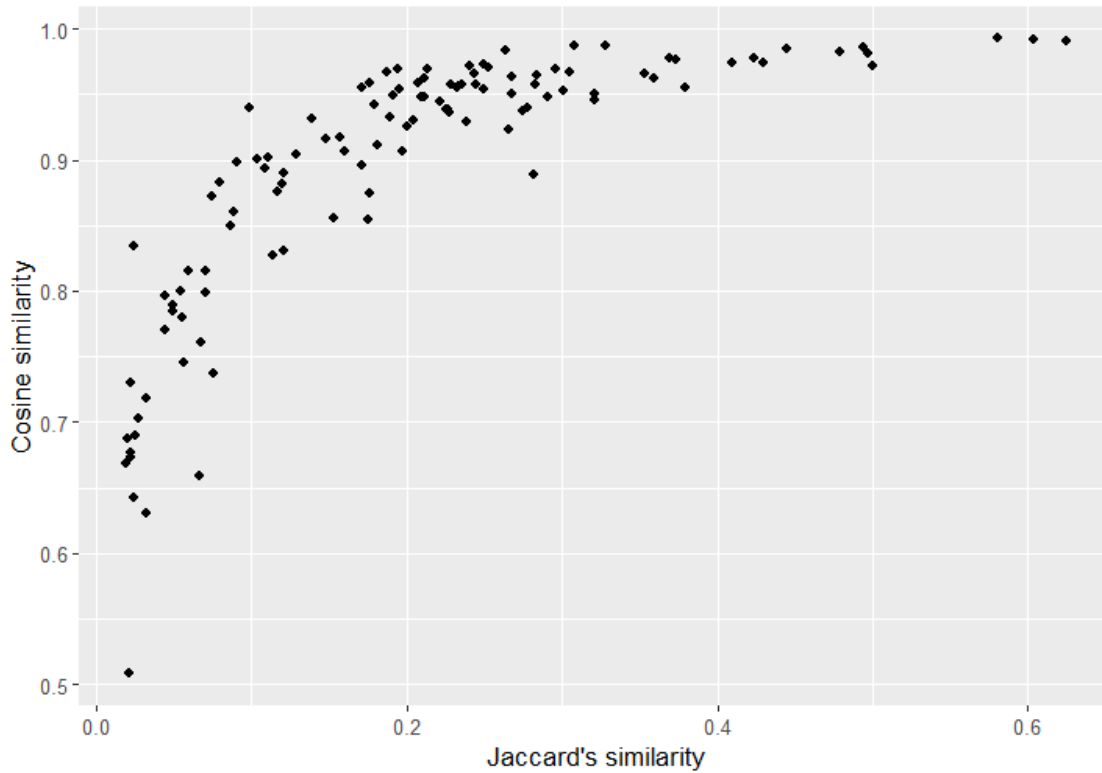

**Figure 5: Scatter plot of Jaccard's and cosine similarity scores.**

In sum, we reject hypothesis 2 since the LLM search does not capture all the keyword results. However, we find that it captures most of them and results are semantically similar to the keyword results.

Hypothesis 3: The most relevant search results (i.e., first 10 results) are different on both search tools since they capture and prioritise different issues.

We next focused just on the immediate search results (defined here as the first ten results presented) since both tools orders results by level of relevancy. Evidence suggests that users tend to focus on the first page of search engines (typically this contains 10 results) and place greater emphasis on the first few results (Silverstein et al. 1999; Pan et al. 2007; Keane et al. 2008). Comparing these results therefore gives a closer representation of how people search for information than compared to looking at all results.

We examined whether the same datasets were present in both top ten results lists. Figure 6 presents the distribution of results. 60% of search terms saw the LLM results containing fewer than half of the keyword search results. For the top 10 results, 13.9% of queries gave the same results in the LLM search compared to the keyword search. 14.8% of queries had none of top 10 results in the keyword search present in the LLM top ten.

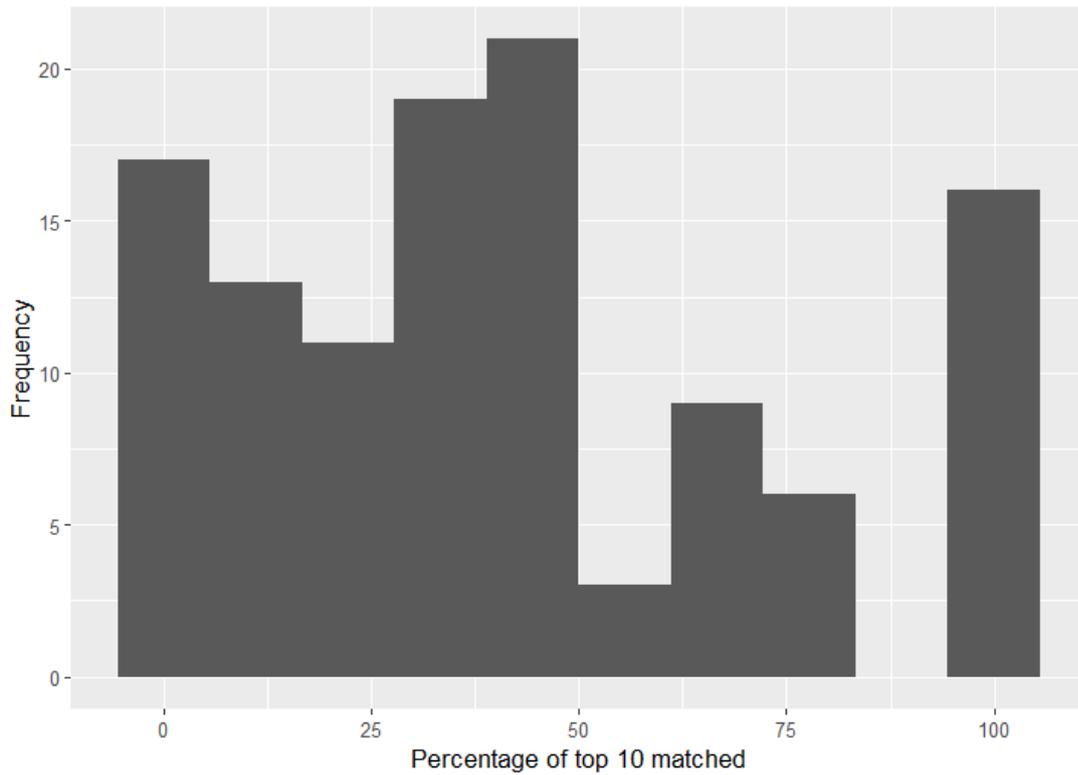

**Figure 6: A histogram of results in the top 10 keyword search results that were present in the LLM search.**

Focusing on the 17 results where there were no results in the top 10 that matched, 10 (58.8%) involved keyword searches that returned only a single dataset. Where we examined queries with at least five keyword search results, the query terms were either general terms (e.g., 'data' or 'maps') or the LLM gave datasets that had some relevance to the topic (for the example of 'employment' see Table 3). If we just consider the first (i.e., top) result returned, 17.4% of queries matched exactly (82.6% did not). 54.8% of queries contained at least one of the top three keyword search results in the top three results returned in the LLM (21.7% had at least two).

**Table 3: Search results returned for the query 'employment'. Note: Only five and four results were returned entirely for the keyword and LLM queries respectively.**

| Keyword tool | LLM tool |
|---|---|
| Index of Multiple Deprivation (IMD) | Output Area Classification (2021) |
| Classification of Workplace Zones (COWZ) | London OAC (2011) |
| London Workplace Zone Classification | Local Data Company - Retail Type or Vacancy Classification |
| MaPS Financial Wellbeing Survey | MIAC Analytics Rental Index |
| MaPS Debt Need Survey | |

The summary statistics for Jaccard's similarity were higher (median 0.21, IQR = 0.14, 0.29) when we computed them for the top 10 results than compared to the full list of search results (21.2% had values >=0.3 and 16.8% less than 0.1). There was little difference for cosine similarity (median 0.94, IQR = 0.89, 0.96), although a higher percentage had values greater than equal to 0.7 (99.1%; 69% were greater than 0.9). The distributions of the two indicators were similar to those presented in Figures 3 and 4, just with a small movement to the right-hand side on Jaccard's similarity reflecting an increase in scores.

We accept hypothesis 3. Although the LLM does not return exactly the same datasets in the most relevant results returned, it does return semantically similar datasets that are relevant.

Hypothesis 4: The LLM semantic search can better handle erroneous, obscure or complex queries than the keyword search.

The final section examines queries where we expect the LLM semantic search tool to perform better. We first start with queries involving terms that have been spelled incorrectly. We expect the LLM tool to interpret spelling mistakes as their patterns match closely to their correct forms – something not possible in the keyword search tool as it requires exact matches. To examine this, we use a case study query of 'retail'. We entered 21 different ways of misspelling retail that could easily happen when typing the query into both search tools.

All the 21 incorrect spellings of 'retail' produced zero results on the keyword search tool. This is because it will not match with the correct spelling and therefore gives no results.

Table 4 presents the results using the LLM semantic search tool. We present the number of queries returned and how similar they were to the keyword search tool results for 'retail'. The results suggest that the LLM tool is quite robust to most incorrect spellings of retail with high values for Jaccard's and cosine similarity scores (median Jaccard's similarity = 0.31, median cosine similarity = 0.98). Indeed, some queries (e.g., 'reetail' and 'rretail') gave all the results from the keyword search for 'retail'. This was due to those search results returning a far larger number of datasets (correlation between number of datasets returned and percent matching those in the keyword results was r = 0.90).

When we clicked on the 'Ask AI' button for top result of all the queries, it did not correct us anytime that we had misspelled the term. Each time it would refer to retail for describing decisions alongside the misspelling.

**Table 4: Comparison of the LLM semantic search performance compared to the keyword search results for the query 'retail'. Note: the keyword search returned 34 datasets.**

| Query | Number of datasets | Keyword datasets matched (%) | Jaccard's similarity | Cosine similarity |
|---|---|---|---|---|
| Correct spelling | | | | |
| retail | 41 | 41.2 | 0.31 | 0.99 |
| Incorrect spellings | | | | |
| Ratail | 70 | 58.8 | 0.35 | 0.98 |
| Reatail | 110 | 82.4 | 0.28 | 0.96 |
| Retal | 15 | 17.6 | 0.21 | 0.95 |
| Retale | 49 | 44.1 | 0.31 | 0.98 |
| Reteal | 49 | 44.1 | 0.32 | 0.97 |
| Reteil | 74 | 61.8 | 0.36 | 0.97 |
| Ritail | 47 | 41.2 | 0.35 | 0.98 |
| Rtail | 69 | 61.8 | 0.37 | 0.98 |
| r3tail | 58 | 50.0 | 0.33 | 0.97 |
| reetail | 244 | 100.0 | 0.31 | 0.97 |
| ret4il | 33 | 26.5 | 0.29 | 0.98 |
| retaail | 55 | 50.0 | 0.35 | 0.98 |
| retai1 | 38 | 32.4 | 0.29 | 0.98 |
| retaiil | 50 | 38.2 | 0.27 | 0.98 |
| retaill | 44 | 41.2 | 0.31 | 0.98 |
| retails | 49 | 41.2 | 0.30 | 0.98 |
| retale | 56 | 64.7 | 0.31 | 0.92 |
| retil | 45 | 41.2 | 0.31 | 0.98 |
| rettail | 51 | 44.1 | 0.31 | 0.97 |
| rretail | 171 | 100.0 | 0.33 | 0.97 |

We further examined the results returned in the lowest similarity scores to explore what datasets were being returned instead. The term 'Retal' produced 15 results of which all results were relevant to retail applications upon inspection (e.g., Geolytix aggregated in-app location dataset, Retail Centre Boundaries and Open Indicators, GambleAware Treatment and Support Survey Data, High Street Retailer - Retail and Consumer Data (2012 - 2017 only)). The same findings were observed when we examined 'ret4il' and 'retai1' which mostly returned relevant retail datasets, but each returned some results that were less relevant (e.g., London OAC (2011), The Ageing in Place Classification, British Population Survey). These datasets included work/employment measures which the LLM felt could be relevant to retail.

If we consider the similarity of the incorrect query spellings to the results returned by the LLM for the correct spelling (Table 5), we see that the incorrect spellings all return results that are close to the LLM result (e.g., median matched datasets = 78.1%, median Jaccard's similarity = 0.60, median cosine similarity = 0.99).

**Table 5: Performance of search results for incorrect spellings compared to the LLM semantic search results for 'retail' (rather than keyword results as in Table 4).**

| Query | Keyword datasets matched (%) | Jaccard's similarity | Cosine similarity |
|---|---|---|---|
| r3tail | 68.3 | 0.47 | 0.98 |
| Ratail | 73.2 | 0.45 | 0.98 |
| Reatail | 100.0 | 0.81 | 0.97 |
| reetail | 100.0 | 0.77 | 0.99 |
| ret4il | 43.9 | 0.44 | 0.99 |
| retaail | 97.6 | 0.79 | 0.99 |
| retai1 | 63.4 | 0.58 | 0.99 |
| retaiil | 75.6 | 0.60 | 0.99 |
| retaill | 90.2 | 0.86 | 0.99 |
| retails | 97.6 | 0.87 | 0.99 |
| Retal | 29.3 | 0.37 | 0.96 |
| retale | 100.0 | 0.51 | 0.93 |
| Retale | 78.0 | 0.65 | 0.99 |
| Reteal | 75.6 | 0.60 | 0.98 |
| Reteil | 100.0 | 0.63 | 0.98 |
| retil | 65.9 | 0.52 | 0.99 |
| rettail | 100.0 | 0.87 | 0.98 |
| Ritail | 51.2 | 0.41 | 0.98 |
| rretail | 100.0 | 0.80 | 0.98 |
| Rtail | 78.0 | 0.49 | 0.99 |

We next considered a case study of query terms that might be obscure to be found in keyword searches. We considered health conditions by searching for terms 'diabetes', 'dementia', and 'lung cancer'. Each term produced no results in the keyword search tool. Entering them into the LLM semantic search gave five, four and six results respectively. Results returned included datasets which contained health outcomes within them that do not have all health conditions listed as keywords (e.g., NHS Hospital Admission Rates by Ethnic Group and other Characteristics, Local morbidity rates of Global Burden of Disease and alcohol-related conditions) and datasets that can provide contextual explanations for disease patterns (e.g., 'Air Pollution Sensors in Liverpool' for lung cancer query). The examples demonstrate how the LLM semantic search query has a better grasp of what a user is searching that could help with data discovery.

The following case study considered whether different results would be returned by the LLM search tool for the exact same query. Inconsistency of results is noted as a common issue when using LLMs, but is not be an issue for keyword based search tools. Again we used an example of 'retail' as the query. Entering this term five times on the LLM semantic search tool produced the same results each time.

Our final case study concerns more complex queries. We trialled a series of queries of increasing complexity in both search tools (Table 6). The keyword search tool struggled to give results when using more than one term. This is something that the LLM semantic search tool can handle better. The LLM semantic search tool have results each time and a larger number of results. The results returned were relevant, with the top 10 results returned always containing retail or footfall datasets (e.g., 'Retail Centre Boundaries and Open Indicators' or 'Retail Centre Boundaries (Previous Versions)' were the top ranked result each time)

**Table 6: Comparison of different queries using the keyword and LLM semantic search tools.**

| Keyword query | Keyword results | LLM query | LLM results |
| --- | --- | --- | --- |
| retail and footfall | 5 | Which retail centres have the highest levels of footfall? | 124 |
| retail and visitors | 0 | Which retail centers attract the largest number of visitors? | 67 |
| shopping and footfall | 0 | Which shopping centres have the highest levels of footfall? | 57 |
| shopping and visitors | 0 | Which shopping centers attract the largest number of visitors? | 60 |

We therefore accept hypothesis 4.